\documentclass [aps,prx,twocolumn,groupeauthor,showpacs,superscriptaddress,longbibliography]{revtex4-1}

\usepackage{amsmath}
\usepackage{amssymb}
\usepackage{bm}

\usepackage{graphicx}
\usepackage{color}

\newcommand{\hloc}{h_{loc}}
\newcommand{\dc}{\delta_{C}}
\newcommand{\tc}{T_{C}}
\newcommand{\thc}{\theta_{C}}

\DeclareMathOperator{\drr}{\frac{\Delta R}{R}}

\begin{document}
\title{Probing the fluctuations of the optical properties in time-resolved spectroscopy}

\author{Francesco Randi}
\affiliation{Department of Physics, Universit\`{a} degli Studi di Trieste, 34127 Trieste, Italy}
\affiliation{Max Planck Institute for the Structure and Dynamics of Matter, 22761 Hamburg, Germany}

\author{Martina Esposito}
\affiliation{Department of Physics, Universit\`{a} degli Studi di Trieste, 34127 Trieste, Italy}

\author{Francesca Giusti}
\affiliation{Department of Physics, Universit\`{a} degli Studi di Trieste, 34127 Trieste, Italy}

\author{Oleg Misochko}
\affiliation{Institute of Solid State Physics, Russian Academy of Sciences, Chernogolovka, Moscow Region 142432, Russia}
\affiliation{Institute of Nanotechnologies of Microelectronics, Russian Academy of Sciences, Moscow 119334, Russia}

\author{Fulvio Parmigiani}
\affiliation{Department of Physics, Universit\`{a} degli Studi di Trieste, 34127 Trieste, Italy}
\affiliation{Sincrotrone Trieste SCpA, 34127 Basovizza, Italy}
\affiliation{International Faculty, Universit\"{a}t zu K\"{o}ln, 50937 K\"{o}ln, Germany}

\author{Daniele Fausti}
\email[Corresponding author: ]{daniele.fausti@elettra.eu}
\affiliation{Department of Physics, Universit\`{a} degli Studi di Trieste, 34127 Trieste, Italy}
\affiliation{Sincrotrone Trieste SCpA, 34127 Basovizza, Italy}

\author{Martin Eckstein}
\email[Corresponding author: ]{martin.eckstein@fau.de}
\affiliation{Max Planck Institute for the Structure and Dynamics of Matter, 22761 Hamburg, Germany}
\affiliation{Department of Physics, University of Erlangen-N\"urnberg, 91058 Erlangen, Germany}

%%\pacs{0000}

\begin{abstract}
We show that, in optical pump-probe experiments on bulk samples, the statistical distribution of the intensity of ultrashort light pulses after the interaction with a nonequilibrium complex material can be used to measure the time-dependent noise of the current in the system. We illustrate the general arguments for a photo-excited Peierls material. The transient noise spectroscopy allows to measure to what extent electronic degrees of freedom dynamically obey the fluctuation-dissipation theorem, and how well they thermalize during the coherent lattice vibrations. The proposed statistical measurement developed here provides a new general framework to retrieve dynamical information on the excited distributions in nonequilibrium experiments which could be extended to other degrees of freedom of magnetic or vibrational origin.
\end{abstract}
\date{\today}
\maketitle

Pump-probe experiments are the prime way to study condensed matter out of its equilibrium state on timescales of femto- and picoseconds. In optical pump-probe experiments, ultrashort pulses are used in pairs. The pump triggers the dynamical response and the probe is used to detect changes in the optical properties of the sample. By and large, the experiments performed to date measure the intensity variation of probe pulses after their interaction with the sample for each pump-probe delay, following the protocol of averaging over many stroboscopically repeated experiments~\cite{Giannetti2016}. Little attention has been given to the accurate measurement of the fluctuation of the intensity of reflected (or transmitted) probe pulses.

In the basic implementation of a ``statistical'' pump-probe set-up, the intensity of every single probe pulse is separately acquired with low-electronic-noise detectors, for every pump-probe delay. This allows to measure both the average of the intensity, which gives the usual pump-probe signal (e.g. the relative variation of the reflectivity $\frac{\Delta R}{R}(t_p)$), and its statistical distribution to all orders. Recent technological advances enabled experiments delivering such full statistical information on intensity fluctuations~\cite{Esposito2015,Tesi,Misochko2001,Riek2017}, thereby providing an experimental benchmark to address the following interesting, yet largely unexplored, question: what is the spectroscopic information carried by the fluctuations of the intensity of ultrashort light pulses reflected (or transmitted)  by complex materials out of equilibrium?

It is known that the photon counting noise in the radiation emitted by nanoscopic emitters contains valuable information on their steady state transport properties~\cite{Beenakker2001,Lebedev2010}, but for pump-probe measurements in bulk samples it is still unclear how to connect the fluctuations of a measurement which is spatially averaged over the optical beam section to microscopic properties of the material. Proof of principle experiments in this direction have shown that, for transparent materials, i.e. materials where a large optical gap freezes the electronic degrees of freedom and an effective photon-phonon Raman coupling is the leading interaction, the statistical fluctuations of the probe intensity can be used to map the temporal evolution of vibrational observables~\cite{Benatti2017,KelvinThesis,Esposito2015,Tesi}. Nevertheless, in the more general setting of pump-probe experiments on complex absorbing materials, where the photo-excitation can trigger a dynamical response in the electronic system, a formalism capable of linking the intensity fluctuations of the probe pulses to the microscopic properties of the material is still unavailable. In this paper we provide such formalism and show how the knowledge of the fluctuations of the optical properties of a system can give access to 
the spectrum of current fluctuations, which is not accessible from the average intensity in a pump-probe experiment.
The general theoretical result will be illustrated with numerical simulations and experimental data for photo-induced coherent lattice vibrations in a charge-density wave system. 

{\em Intensity fluctuations ---} 
A pump-probe experiment measures the number $m$ of photo-counts in a detector, during a detection time-window which is very large compared to the probe-pulse duration (in general, larger by at least three orders of magnitude). In the statistical pump-probe experiment one obtains in addition the variance $\Delta m^2 = \sigma[I]+ \langle m \rangle$, which can be written as the sum of a shot-noise contribution $\langle m\rangle$ proportional to the intensity itself, and a sample-dependent contribution $\sigma[I]$. Because the induced fields at the detector are linearly related to their sources, i.e., the current density $j$ in the sample, the probability distribution $\mathcal{P}(m)$ can be written in terms of the time-ordered current correlation functions \cite{Fleischhauer1998}. In the following we show that the nonequilibrium current fluctuations generated by the pump give rise to a contribution $\sigma_\text{bulk}$ to  $\sigma[I]$ which can be identified by the  statistical pump-probe experiment on a bulk sample, because it is of leading order in the sample volume. This contribution can be understood as  the interference term $\langle  |\int dt E_+(t) \varepsilon(t) |^2\rangle_\text{noise}$ of the (conventionally averaged) reflected probe field $E(t)$ at the detector ($E_+$ is the positive frequency component $\sim e^{-i\omega t}$), \and a field $\varepsilon(t)$ which is generated at the detector by a classical noise current $\eta$ with a Gaussian distribution and variance given by the current fluctuations in the sample, $\langle \eta(1) \eta(1')\rangle_\text{noise} = \langle \delta j(1) \delta j(1') \rangle$. (Here and in the following, $1\equiv (\bm r_1,t_1)$ denote space-time points).  Since Maxwell's equations relate the current density $j(1)$ at point $1$ to the induced fields $E_\text{ind}(2)=\int d1 g(2,1) j(1)$ at a point $2$ via a linear kernel $g(2,1)$, the above statement is equivalent to
\begin{align}
&
\sigma_\text{bulk}
=\epsilon^2\!\!
\int \!\!
dt dt'
E_+(t) E_-(t')
\langle
\delta j(t,\bm R) \delta j(t',\bm R) 
\rangle_\text{ret},
\label{eqI33}
\end{align}
where 
$\langle j(1) ...\rangle_\text{ret} \equiv \int d\bar 1g(1,\bar 1) \langle j(\bar 1) ...\rangle$ is the expectation value of the currents propagated to space-time point $1$, $\bm R$ is the location of the detector, and $\epsilon$ is proportional to the detector volume and efficiency. To lowest order in the probe amplitude, the correlation function in Eq.~\eqref{eqI33} can be evaluated without the effect of the probe. Hence the probe pulse can be chosen to project out the transient current noise spectrum of the nonequilibrium state generated by the pump at suitable times $t,t'$ (or as a function of time and frequency).
%%%%%

To prove Eq.~\eqref{eqI33} we start from Ref.~\cite{Fleischhauer1998}, which gives the factorial moments $I_1=\langle m\rangle $, $I_2=\langle m (m-1)\rangle $ of the photo-count distribution $\mathcal{P}$ in terms of the sources,
\begin{align}
I_n=&
\int\! d1d1' \cdots dndn' \,\,\mathcal{D}_{1,1'}\cdots \mathcal{D}_{n,n'}
\nonumber \\
&\times \,\,\,
\langle
T_{\bar \tau}[j(1)\cdots j(n)]T_\tau[j(1')\cdots j(n')]
\rangle_\text{ret}.
\label{eqI2}
\end{align}
Here $T_{\tau}$ ($T_{\bar \tau}$) is the (anti) time-ordering operator acting on the sources, and $\mathcal{D}_{1,1'}$ is the detector response function. For a small and wide-band detector at site $\bm R$, which absorbs at all positive frequencies, the latter is simply  $\mathcal{D}_{1,1'}=\epsilon \delta(\bm r_1-\bm R)\delta(\bm r_1'-\bm R) \int_{0}^\infty \! d\omega  \,e^{-i\omega (t_1-t_1')}$. Differently from the standard quantum theory of photo-detection \cite{Mandel1958,Kelley1964},  Eq.~\eqref{eqI2} is not obtained within the rotating wave approximation, and is thus applicable down to ultra-short (possibly single-cycle) pulses \cite{Fleischhauer1998}. The variance of the photo-count is $\Delta m^2 = \langle m^2 \rangle-\langle m \rangle^2 = I_2-I_1^2 + \langle m \rangle$, so that $\sigma[I]=I_2-I_1^2$. To evaluate $\sigma[I]$, we shift the operators $j(t)=\langle j(t) \rangle+\delta j(t)$, and expand $\sigma[I]$ in the fluctuations. In a bulk sample, terms like $\int d^3{\bm r_1} d^3{\bm r_2} \cdots \langle \delta j(\bm r_1) \delta j(\bm r_2) \cdots\rangle$ are proportional to the volume $V$, since they are obtained as the derivative of the extensive free energy with respect to an external vector potential. Hence, the dominant contribution ($\sim V^3$)  to $\sigma[I]$  is given by the second order terms like $ \langle j(1)  \rangle \langle \delta j(2) \delta j(3) \rangle  \langle j(4) \rangle $ (first and zeroth order terms vanish by construction). Since $\langle j(\bm R, t)\rangle_\text{ret} \equiv E(t)$ is just the classical reflected field at the detector, and the contraction with the detector function projects out it's positive and negative frequency parts,  $\int d1' \mathcal{D}_{1,1'} \langle j(1')\rangle_\text{ret} \equiv E_{+}(t_1)$ and $\int d1' \mathcal{D}_{1',1} \langle j(1')\rangle_\text{ret} \equiv E_{-}(t_1)$, one arrives at Eq.~\eqref{eqI33} 
%\cite{footnote1,Supplemental}.
\cite{footnote1} (see appendix).

While Eq.~\eqref{eqI33} holds for an arbitrary geometry, to obtain specific results we adopt the standard setup and neglect propagation effects in the sample, so that $E_\text{ind}(t,\bm R)=-\dot A(t)$ with $A(t)\propto \int_\text{sample} d^3\bm r \,j(\bm r,t_\text{ret})$, where $t_\text{ret}=t-\Delta t$ is  shifted by the propagation time $\Delta t$ from the sample to the detector. This gives
\begin{align}
\label{eqI33-a}
\sigma_\text{bulk}
\propto
\epsilon^2 \int  dt dt' \,\langle \delta  J(t_\text{ret}) \delta  J(t'_\text{ret})  \rangle  \dot  E_+(t) \dot  E_-(t'),
\end{align}
where $J$ is the current integrated over the sample volume.

When the system is in a thermal equilibrium state, the current fluctuations   $C(t,t')=\langle   J(t')  J(t)  \rangle$ are related to the response function $\chi^R(t,t')=i\theta(t-t') \langle [J(t), J(t')] \rangle$ of time-resolved optical spectroscopy \cite{Eckstein2008} by a fluctuation-dissipation theorem (FDT)~\cite{Kubo1957}, 
\begin{equation}\label{eq:fluctuationdissipation}
C(t_p,\omega) =
2\, b(\omega,\beta)\, \text{Im} \chi^R(t_p,\omega),
\end{equation}
where $f(t_p,\omega)= \frac{1}{2\pi}\int ds\,\,e^{i\omega s} f(t_p+s/2,t_p-s/2)$
is the Fourier transform with respect to difference time,
and $b(\omega,\beta)=(e^{\beta \omega} - 1 )^{-1}$ is the Bose function with inverse temperature $\beta$. 
In an equilibrium state, we take $ \dot  E_\pm(t)\sim e^{\mp i\omega t} $, so that $\sigma_\text{bulk}\propto C(\omega)$, which is therefore negligible  at room temperature for probe photon energies in the visible and infrared spectral range. Only the shot noise is then relevant, as it is usually assumed in quantum optics for the reflection of light by a mirror. Note that in our language, contributions to $\sigma[I]$ on top of $\sigma_\text{bulk}$ which are subleading in the volume $(\sim V^2)$, can be nonzero in equilibrium, scale like the intensity $E^2$ and add to the shot noise of the light coming from an ``imperfect reflector''.  
When the sample is out of equilibrium, instead, $\langle  J(t)  J(t') \rangle$ can significantly contribute to the total fluctuation of the intensity, and its  measurement can therefore give information on the nonequilibrium state.

\begin{figure}[t]
\centering
\includegraphics[width=1\linewidth]{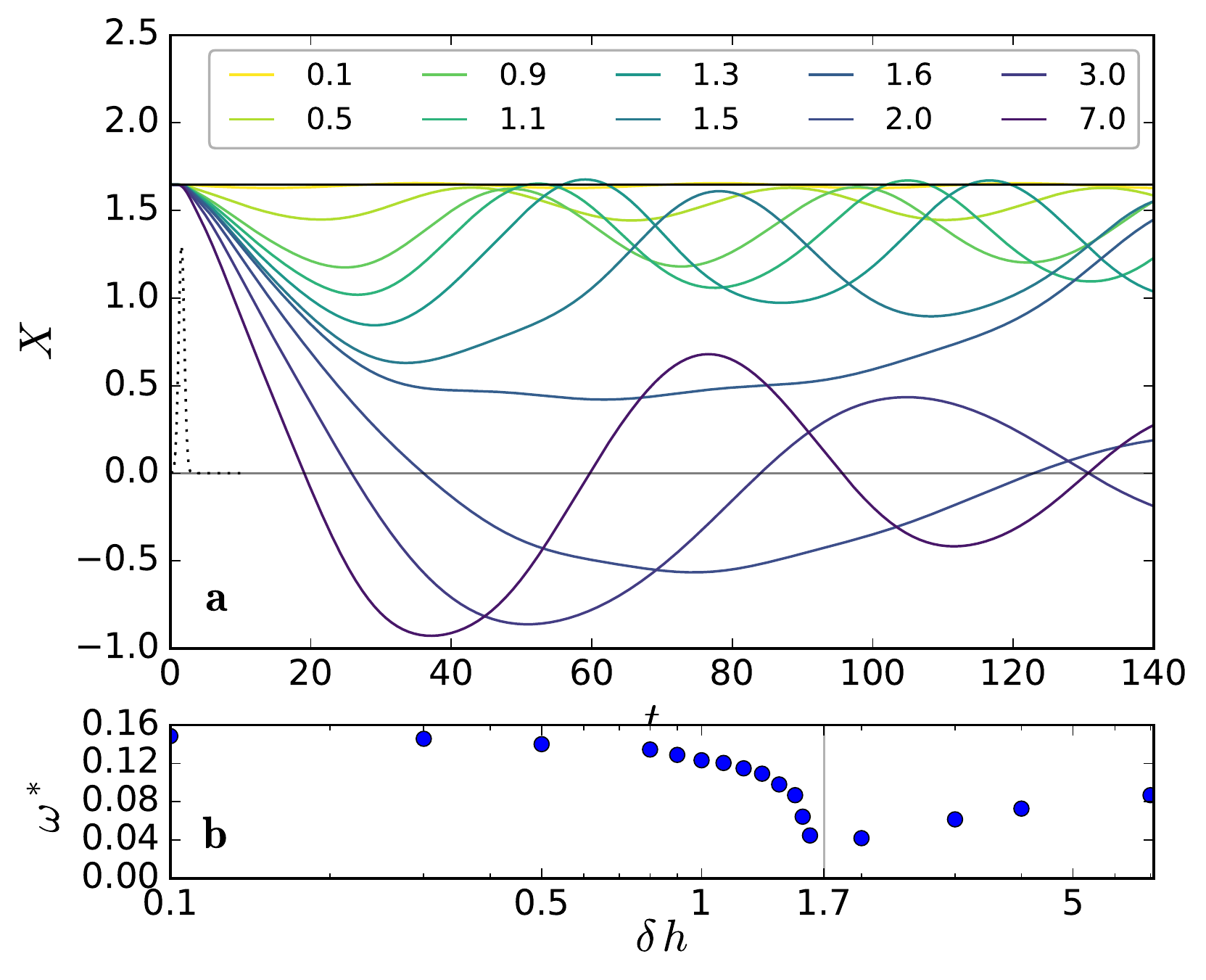}
\caption{
\textbf{
a)} Atomic displacement $\langle X \rangle$ as a function of time $t$, for different excitation strength $\delta h$ above and below the critical excitation $\delta h_c = 1.7$ for melting of the CDW. The dotted line shows the hopping modulation  (in arb.~units) used to impulsively stimulate the system.
\textbf{b)} Oscillation frequency $\omega^*$ as a function of $\delta h$. The grey vertical line indicates $\delta h_c$, where a critical slow-down is observed.
}
\label{fig:X}
\end{figure}

To illustrate such approach both from the theoretical and experimental point of view, we concentrate on a Peierls charge-density wave (CDW) system. A prototypical realization is bismuth~\cite{Peierls1991}, where an impulsive excitation generates coherent vibrations of the $A_{1g}$ phonon, which modulates the out-of-equilibrium reflectivity \cite{Misochko2001}. A minimal theoretical model for the Peierls system is the Holstein model, which describes free electrons coupled to a dispersion-less vibrational mode (Einstein phonon). At half filling, the Hamiltonian is
\begin{equation}\label{eq:HolsteinHamiltonian} 
H = -t_0\sum\limits_{\langle ij \rangle \sigma } c_{i\sigma}^\dagger c_{j\sigma} + \omega _0\sum_j b^\dagger_j b_j + g
\sqrt{2}
 \sum\limits_{j} X_j (
n_{j}-1),
\end{equation}
where $t_0$ is the nearest neighbor hopping, $c^\dagger_{j\sigma}$ creates an electron at lattice site $j$ (spin $\sigma$), $b_j$ and $b^\dagger_j$ are the bosonic operators of a phonon mode at frequency $\omega_0$, and $g$ is the coupling constant between the electronic density $n_{j}=n_{j\uparrow}+n_{j\downarrow}$ and the phonon coordinate $X_j=(b^\dagger_j+b_j)$. Below the critical temperature, the system is in an insulating and symmetry broken phase, with a staggered charge disproportionation on neighboring sites and a gap at the Fermi level. We have numerically studied the out-of-equilibrium dynamics of the Holstein model using the non-equilibrium dynamical mean-field theory~\cite{NonEqDMFT} within the self-consistent Migdal approximation~\cite{Murakami2015}, on a bipartite 
%lattice~\cite{Supplemental}. 
lattice~(see appendix).
We choose $g = 0.34$, the inverse temperature $\beta = 25$, and $\omega_0 = 0.2$. The free bandwidth $W=4$ sets the energy and  time-scale ($\hbar=1$). The system is brought out of its equilibrium state by a short modulation of the hopping $t_0\to t_0(1+\delta h\, e^{-(t-0.16)^2/0.32})$, see dotted line in Fig~\ref{fig:X}a, which impulsively creates a conduction band electron population, analogous to a pump pulse impinging on a sample. (The precise excitation mechanism is not important, since we focus on the subsequent relaxation dynamics, which depends mainly on the excitation density controlled by $\delta h$.) 

\begin{figure}[t]
\centering
\includegraphics[width=1.0\linewidth]{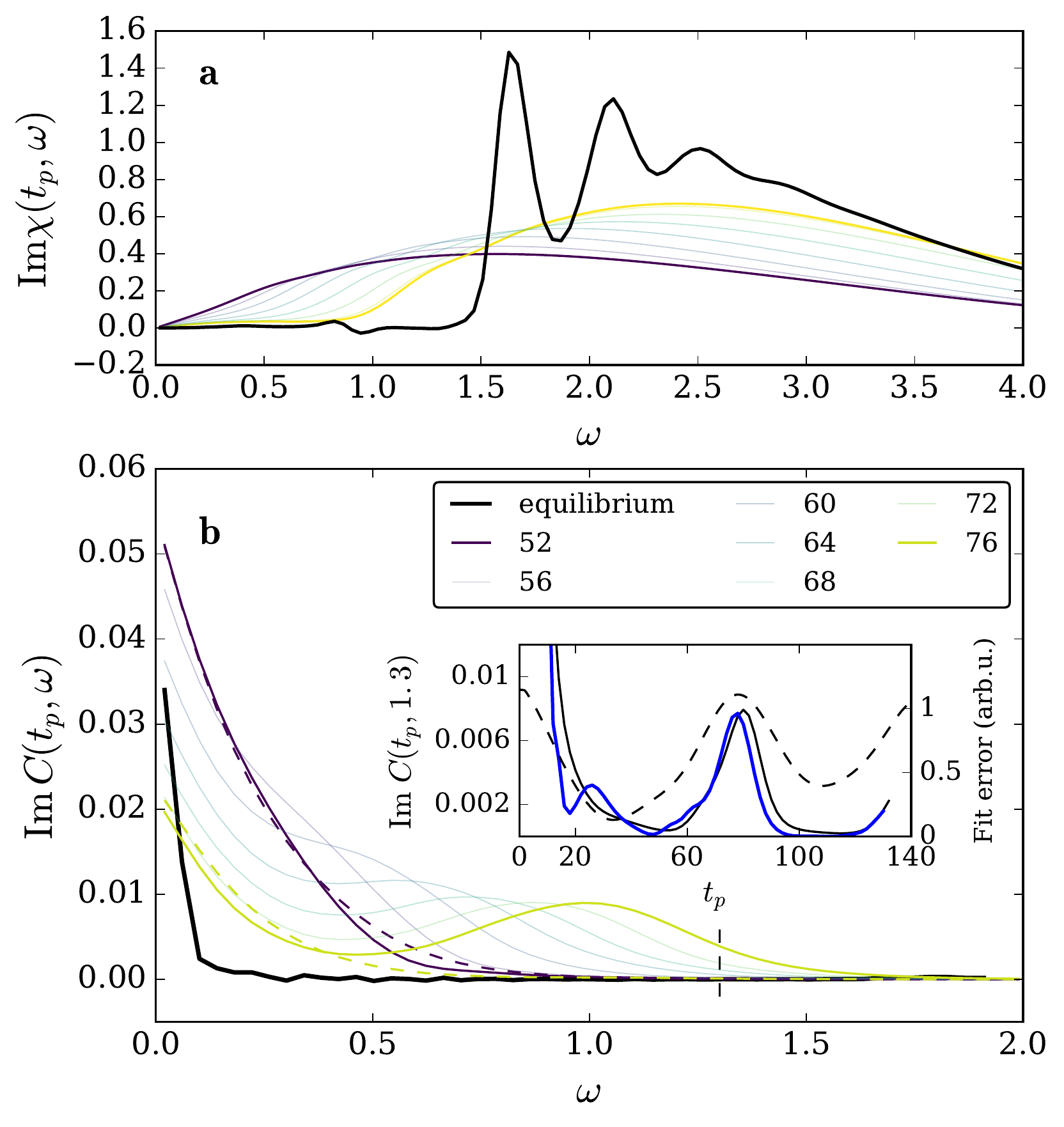}
\caption{
Optical susceptibility $\chi^R(t_p,\omega)$ \textbf{(a)} and current fluctuations \textbf{(b)}  at equilibrium (black), and for different $t_p$. ($\delta h=1.5$). Dashed lines in \textbf{(b)} correspond to the right-hand side of the FDT
 \eqref{eq:fluctuationdissipation}, with 
$\beta_f=7.1$ and $6.3$ for $t_p=52,76$, respectively. \textbf{Inset)} 
Violation of the FDT (blue line, see text), $\langle X(t)\rangle$ (dashed line), and reflectivity fluctuations $\sigma_\text{bulk}$ [Eq.~\eqref{eqI33-a}] at probe frequency $\omega=1.3$ as a function of $t_p$.
}
\label{fig:jj}
\end{figure}

Before discussing the nonequilibrium current fluctuations, we verify that this model captures the main features which are known from experiments on CDW systems \cite{Misochko2001}, i.e., a coherent oscillation which is increasingly damped and softened for larger excitation density. A good observable to study this dynamics is the displacement $\langle X \rangle$ of the atoms from  the translationally invariant positions of the high-temperature phase ($\langle X \rangle$ takes opposite values on the two sub-lattices of the bipartite lattice). As shown in Fig.~\ref{fig:X}a, for $\delta h$ below a critical value $\delta h_c\simeq 1.7$, $\langle X(t) \rangle$ coherently oscillates around a new non-zero position. ($\delta h_c\simeq 1.7$ corresponds to an pump-induced conduction band population $n_{ex}=0.05$.) As the excitation density increases, the oscillations are damped more quickly, and the frequency $\omega^*$ becomes lower, i.e. the vibrational mode is softened (see Fig.~\ref{fig:X}b). For $\delta h > \delta h_c$, the melting of the  CDW phase is induced, with oscillations around the undistorted atomic position $\langle X \rangle=0$ (Fig.~\ref{fig:X}a).  While the melting of the CDW phase has been observed in experiment~\cite{Teitelbaum2016}, we will report measurements below the threshold and therefore focus on the regime $\delta h < \delta h_c$ for the following theoretical analysis.

Besides the distortion, we calculate the optical susceptibility $\chi^R(t_p,\omega)$ and the fluctuations $C(t_p,\omega)$, as defined above Eq.~\eqref{eq:fluctuationdissipation}. At equilibrium, the optical susceptibility $\chi^R(\omega)$ of the CDW phase displays a gapped optical band (Fig.~\ref{fig:jj}a).  After the excitation, the out-of-equilibrium $\text{Im} \chi(t_p, \omega)$ oscillates between a gapped spectrum, which is partially filled by intraband transitions due to the excited conduction band electron population (Fig.~\ref{fig:jj}a, $t_p=76$), and an almost gapless spectrum when the atoms are closest to their translationally invariant positions $\langle X \rangle=0$ (Fig.~\ref{fig:jj}a, $t_p=52$). The spectrum $C(t_p,\omega)$ shows that current fluctuations exist only at low frequencies (black line) in equilibrium, but  extend to high frequencies when  a non-thermal population is created. 

A common assumption is that the electrons in a solid, after an intense photo-excitation, can be considered in an effectively thermal state. In systems with strongly coupled electrons and phonons, this assumption need not hold, in particular when phonons itself are far from equilibrium. Theoretically, a check of the FDT
\eqref{eq:fluctuationdissipation}  provides a natural way to access the thermal nature of a nonequilibrium state~\cite{Kogoj2016,NonEqDMFT}. The dashed lines in Fig.~\ref{fig:jj} show the right-hand side of Eq.~\eqref{eq:fluctuationdissipation} for two representative pump-probe delays ($t_p=52,76$), with a best-fit inverse temperature $\beta_f=7.1$ and $6.3$, respectively.  The most pronounced violation of the fluctuation dissipation relation is visible after one full oscillation (yellow curve), i.e. when $\langle X(t) \rangle$ is maximum (see also the blue line in the inset of Fig.~\ref{fig:jj}b, which shows a fit error between $C(t_p,\omega)$ and the right-hand side of Eq.~\eqref{eq:fluctuationdissipation}). One can understand this behavior as follows: Initially, a non-thermal population is created at the top of the lower band and at the bottom of the upper band of the system. As the atoms move closest to $\langle X \rangle=0$ and the gap is the smallest, electrons thermalize and the occupation of the bands closely resembles the Fermi-Dirac distribution, with a high electronic temperature.  But when the gap revives after one period, a nonthermal distribution far from the Fermi energy is restored, because reopening of the gap itself prevents electronic thermalization along with the evolution of $\langle X\rangle$.

\begin{figure}[t]
\centering
\includegraphics[width=\columnwidth,height=0.6\columnwidth]{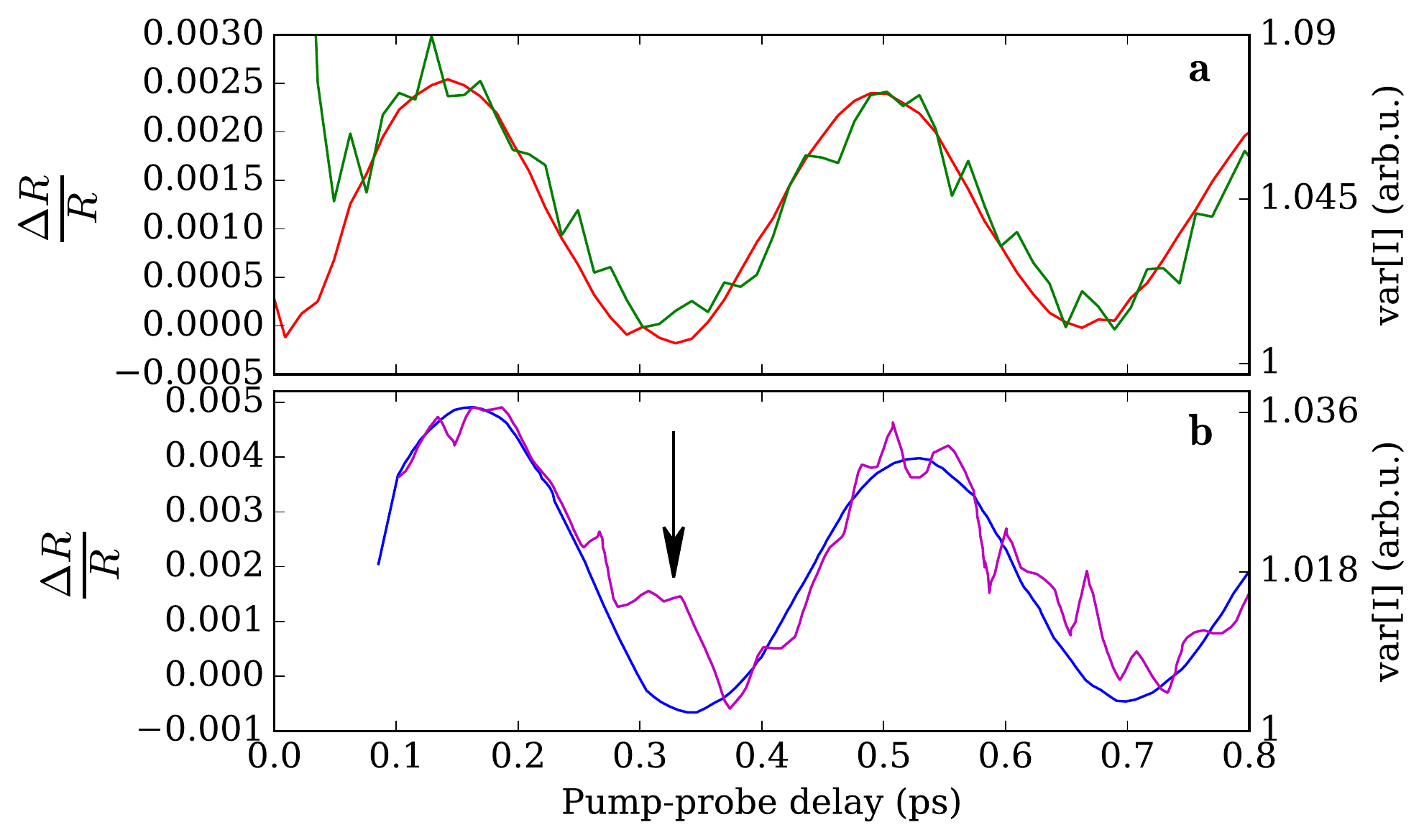}
\caption{
Measured reflectivity change $\drr$ and variance $\Delta m^2$ in a bismuth crystal after optical excitation at a lower \textbf{(a)} and higher \textbf{(b)} excitation intensity. The curves $\Delta m^2$ are rescaled to obtain a best fit to $\drr$. The arrow points at enhanced fluctuations around the first oscillation period.
}
\label{fig:Bi}
\end{figure}

The measurement of the fluctuations of the intensity of the probe pulses now allows to obtain the fluctuations of the current in the sample experimentally. In the present case, the nonthermal nature of the state after one oscillation period can be detected by probing the fluctuations at a specific high frequency.  We evaluate Eq.~\eqref{eqI33-a} assuming a generic pulse $\dot E(t)\propto\cos(\omega_p (t-t_p))e^{-(t-t_p)^2/\delta t^2}$ with frequency $\omega_p=1.3$ larger than the gap, probe time $t_p$, and probe pulse duration $\delta t=2.4$ \cite{footnote02} The result, shown by the black curve in the inset of Fig.~\ref{fig:jj}b, closely resemble the error in the fit of the  FDT. These numerical predictions can be compared to experimental results obtained in experiments on Bi single crystals with the statistical pump-probe 
set-up~\cite{Esposito2015}.  
The system is excited and probed at a photon energy of 1.5 eV, with a probe duration of 80 fs. Figures~\ref{fig:Bi}a and b show the measured variance $\Delta m$ and $\drr$ for two different fluences. The pump generates a coherent amplitude mode oscillation, which is seen in $\drr\propto \langle X\rangle$. Also $\Delta m$ mainly follows this oscillation. This is expected because in addition to the contribution $\sigma_\text{bulk}$ from the nonequilibrium current fluctuations, $\Delta m$ contains shot noise proportional to the intensity itself. Moreover, there is a slight variation with $\drr$, because the measurement of $\Delta m$ is calibrated to the equilibrium reflectivity of the sample. These contributions to $\Delta m$  are quantitatively unknown, but they are proportional to $\drr$, and one can clearly see that a rescaled $\drr$ cannot describe the time-dependent fluctuations $\Delta m$. In particular, we find an enhancement of the fluctuations at early times, and,  at higher fluence,  after one oscillation period, i.e., around the time when the atoms revive to their most distorted positions. This is well in agreement with the theoretical prediction for the fluctuation contribution $\sigma_\text{bulk}$ which comes from the current noise. The experimental data therefore show the that electrons have not thermalized at each time during the early CDW oscillations of the system.  

In conclusion, in this letter we have derived an expression for the fluctuations of the intensity of light reflected by a sample in an ultra-fast pump-probe experiment. We have shown that fluctuations of the reflectivity in addition to shot noise contributions can measure the nonequilibrium current noise spectrum as a function of time and frequency on the femtosecond timescale. In a most straightforward way, current fluctuations can reveal the nonequilibrium state of the sample, such as the effective temperature of the relevant degrees of freedom, or whether the system can be considered in a quasi-thermal state at all. We illustrated our general results for the case of coherent vibrations in solids, both via numerical calculations and statistical pump-probe experiments on bismuth single crystals. We find current fluctuations at high energies, which reveals that nonthermalized electronic distributions emerge as the CDW gap oscillates. An intriguing future direction would be to also characterize the quantum state of the low energy mode. While the theoretical simulations predicts that is does not only coherently oscillate but also becomes squeezed,
% \cite{Supplemental}, 
this is only indirectly reflected in the probed response at high energies. In general, the results of this manuscript show that measuring the fluctuations in the reflectivity opens up an independent way to characterize the nonequilibrium dynamics in bulk solids, with intriguing future applications in the characterization of transient superconducting or nonthermal symmetry broken states.

\acknowledgements
We thank Federico Cilento for the support in the setup of the experiments and useful discussions. Further we thank Arunangshu Debnath, Fabio Benatti, Roberto Floreanini, Kelvin Titimbo and Klaus Zimmerman for the insightful discussion. F.R. was supported by the Friuli Venezia Giulia Region (European Social Fund, operative program 2007/2013). D.F aknowledges the support by the European Commission through the ERCStG2015, INCEPT, grant agreement number 677488. M.E. acknowledges support by the DFG within the Sonderforschungsbereich 925 (project B4).

%%%%%%%%%%%%%%%%%%%%%%%%%%%%%%%%%%%%%%%%%%%%%%%%%%%%%%%%%
%%%%%%%%%%%%%%%%%%%%%%%%%%%%%%%%%%%%%%%%%%%%%%%%%%%%%%%%%

\begin{appendix}

\section{Nonequilibrium DMFT for the Holstein model}\label{sec:Holstein:DMFT:SetUp}

In this section we present the nonequilibrium DMFT setup for the Holstein model. Apart from the symmetry breaking, the formalism is analogous to what has been explained in Ref.~\cite{Murakami2015}. We therefore only state the equations, and do not provide a detailed derivation.

In DMFT, the Holstein model [see Eq. (5) of the main text] is mapped to a set of Anderson-Holstein impurity models (one for each inequivalent lattice site), with action
\begin{align}
S
=
&-i
\sum_{\sigma}
\int_C dt\, \big[\sqrt{2}gX\,(c_\sigma^\dagger c_\sigma-\tfrac12) + \frac{\omega_0}{2}(X^2+P^2)\big]
\nonumber
\\
&-i
\sum_{\sigma}
\int_C dt_1 dt_2 \,c_\sigma^\dagger(t_1) \Delta(t_1,t_2) c_\sigma(t_2)
\label{Simp}
\end{align}
on the Keldysh time-contour $C$. (For an introduction to nonequilibrium DMFT and to the Keldysh formalism, see Ref.~\cite{NonEqDMFT}). In Eq.~\eqref{Simp}, the first term is the local part of the lattice Hamiltonian, which involves the coupling of the electrons at the impurity site to the coordinate $X=(b^\dagger+b)/\sqrt{2}$ of the local oscillator, and  $\Delta(t_1,t_2)$ is the hybridization function, which is determined self-consistently below. 

The impurity model is solved using the self-consistent Migdal approximation \cite{Murakami2015}, where also the vibrational mode evolves as a consequence of the interaction with the electrons.  In the symmetry broken phase, the coordinate $X$ acquires a nonzero expectation value. The expectation value is determined by the exact equation of motion  $\frac{d^2}{dt^2}\langle X(t)\rangle = -\omega_0^2\langle X(t)\rangle + F(t)$, with the time-dependent force $F(t)$
\begin{equation}\label{eq:Holstein:DMFT:Xt}
F(t) = \sqrt{2} g\sum_\sigma \big( \langle c_\sigma^\dagger (t) c_\sigma(t) \rangle - 0.5 \big).
\end{equation}
In turn, there is a time-local (Hartree) contribution to the electronic self-energy, i.e.,  a self-consistent on-site potential,
\begin{equation}\label{eq:Holstein:DMFT:hloc}
\hloc (t) = - \sqrt{2} g \langle X(t) \rangle.
\end{equation}
Furthermore, we include the leading order self-consistent diagrammatic corrections in the expansion in 
terms of the fluctuations $\tilde X = X - \langle X(t) \rangle$. The second-order electronic self-energy is
\begin{equation}\label{eq:Holstein:DMFT:SelfEnergy}
\Sigma(t,t') = i g^2 G(t,t') D(t,t'),
\end{equation}
where
\begin{equation}
D(t,t') = -2i \langle\tc \tilde X(t) \tilde X(t') \rangle.
\end{equation}
(We consider the spin-symmetric phase and omit spin indices $\Sigma_\sigma$ and $G_\sigma$.) 
With this, the Dyson equation for the electronic Green's function reads
\begin{equation}\label{eq:Holstein:DMFT:GreensFunction}\begin{split}
\big( i\partial_t &+ \mu - \hloc\big)G(t,t') - \\
&\big(\Delta(t,t') + \Sigma(t,t')\big)*G(t,t') = \dc(t,t').
\end{split}\end{equation}
To include the back-action of the electrons on the phonons on the same diagrammatic level,  we include the phonon self-energy (polarization operator)
\begin{equation}
P(t,t') = -2i g^2 G(t,t') G(t',t),
\end{equation}
and solve the phonon Dyson equation in the form
 \begin{equation}\label{eq:Holstein:DMFT:Dttp}
\big(1 - D_0(t,t') * P(t,t') \big) * D(t,t') = D_0(t,t').
\end{equation}
Here $D_0(t,t')$ is the non-interacting phonon propagator,
\begin{align}
D_0(t,t') =&
-i
 \big[2 \cos(\omega_0(t-t')b_\beta+
\thc(t',t) e^{i\omega_0(t-t')}
\nonumber
\\
&+ \thc(t,t') e^{-i\omega_0 (t-t')} 
 \big],
\end{align}
where 
$b_\beta=1/(e^{\beta\omega_0}-1)$
is the Bose function.

In the present case of a two-sublattice symmetry broken phase, we have two inequivalent impurity models \eqref{Simp}, which represent sites on the $a$ and $b$ sublattice, i.e., all quantities, $G$, $\Delta$, $\hloc$, $\langle X (t)\rangle, $ $\Sigma$, $P$, will additionally depend on the sublattice $a,b$. For the particle-hole symmetric case, we have $\langle X(t)\rangle_a=-\langle X(t)\rangle _b$. We use a bipartitle lattice with a semielliptic density of states, in which the DMFT self-consistency is given by \cite{NonEqDMFT}
\begin{align}
\Delta_a(t,t')=v(t)G_b(t,t')v(t'),\\
\Delta_b(t,t')=v(t)G_a(t,t')v(t'),
\end{align}
where $v(t)$ is the time-dependent profile of the hopping amplitude. This closes DMFT equations.

{\em Optical conductivity:} The main quantity of interest in this work is the current correlation function $C(t,t')=\langle j(t) j(t')\rangle$, and the 
optical susceptibility $\delta\langle j (t)\rangle / A(t')$, which characterizes the long-wavelength current ($q\rightarrow 0$) in response to an applied time-dependent vector potential $A(t)$.  Since the current operator is given by $j=-\delta H/\delta A$, the latter response function is given by $\chi^{R}(t,t')=i\theta(t,t') \langle[j(t),j(t')] \rangle$. (For simplicity of notation, we are omitting cartesian components $x,y,z$.)

The current-current correlation function is obtained from the lattice Green's function by direct generalization of the expressions presented in Ref.~\cite{Eckstein2008}. Both response and correlation function are obtained from the contour-ordered current-current correlation function, 
\begin{align}
\label{jj}
\chi(t,t')=
i\langle T_C j(t) j(t')\rangle.
\end{align}
which is defined as the response of the current to an arbitrary variation of the vector potential along the Keldysh contour, 
\begin{align}
\label{dlda}
\delta \langle j(t)\rangle = \int_C d\bar t\,  \chi(t,\bar t) \delta A(\bar t),
\end{align}
omitting a diamagnetic contribution which is time-local and thus irrelevant for the discussion of the dynamic properties discussed in this paper. Current fluctuations are given by the greater and lesser component, $\chi^>(t,t') \equiv \chi(t_-,t'_+)= i\langle j(t) j(t')\rangle$ and $\chi^<(t,t') \equiv \chi(t_+,t'_-)= i\langle j(t') j(t)\rangle$ ($t_{\pm}$ is on the upper/lower branch of the Keldysh contour), and $\chi^{R}(t,t') = \theta(t,t')(\chi^>(t,t')-\chi^<(t,t')$.

In the symmetry broken phase, the lattice has a unit cell with two sites $a,b$, and a reduced Brilluoin zone (RBZ).  We introduce the spinor,
\begin{align}
\hat \psi_k= \begin{pmatrix} c_{k,a} \\ c_{k,b}\end{pmatrix},
\end{align}
and the momentum-dependent Green's function $G_k$ then becomes a $2\times2$ matrix,
$\hat G_{k}(t,t') = -i\langle T_C \hat \psi_k(t)\hat \psi_k^\dagger (t')\rangle$.
The hopping term takes the form $H_{hop} = \sum_{k} \hat \psi_{k}^\dagger \epsilon_{k-A} \hat \sigma_1 \hat \psi_k$, with the Pauli matrix $\hat \sigma_{1}$;  $\sum_{k}$ is a sum over the reduced Brillouin zone (RBZ). The vector potential is added by the Peierls substitution $\epsilon_k\to\epsilon_{k-A}$, so that the current
$j=-\delta H/\delta A$
 is given by
\begin{align}
\langle j(t)\rangle
=
-2i\sum_{k\in RBZ}\text{tr} \big[ v_{k-A} \hat \sigma_1 \hat G_k(t_+,t_-)\big],
\end{align}
where $v_k=\partial_k \epsilon_k$ is the band velocity, and the factor $2$ is for spin. 
Following Ref.~\cite{Eckstein2008}, we take the variation $\delta A(t')$, 
using that vertex corrections to the current correlation function vanish in DMFT. 
This gives the susceptibility (evaluated at $A=0$),
\begin{align}
&
\chi(t,t')=2i\sum_{k\in RBZ}  v_{k}^2 \,\text{tr}\big[ \hat \sigma_1 \hat G_k(t_+,t')\hat \sigma_1G_k(t',t_-)\big],
\end{align}
which is the usual bubble diagram of the Green's functions.

In DMFT (and when we consider only a modulation of the hopping amplitude, as in the manuscript), $\hat G_k$ depends on $k$ only via the dispersion $\epsilon_k$, i.e., $\hat G_{k}(t,t')\equiv \hat G_{\epsilon_k}(t,t')$. In the particle-hole symmetric case on a bipartite lattice, the RBZ corresponds to positive values of $\epsilon_k$.  The momentum sum can then be represented by integrals 
\begin{align}
&\sum_{k\in RBZ} f(\epsilon_k) =\int_0^\infty \rho(\epsilon) f(\epsilon),\\
&\sum_{k\in RBZ} v_k^2 f(\epsilon_k) = \int_0^\infty D(\epsilon) f(\epsilon),
\end{align}
where $\rho$ and $D$ depend on the lattice. We assume a semi-elliptic density of states $\rho(\epsilon)=\sqrt{4-\epsilon^2}$, and the corresponding form for $D(\epsilon)$ as defined in Ref.~\cite{Eckstein2008}.  The lattice Green's function is evaluated on a grid of momentum points, solving the Dyson equation
$\hat G_{\epsilon_k}(t,t') = (i\partial_t + \mu - \hat h(t) - \hat \Sigma(t,t'))^{-1}$, where $\hat h$ and $\hat \Sigma$ in the \{a,b\} basis  are
\begin{align}
\hat h(t) &=
\begin{pmatrix}
g X_a(t) && \epsilon_k (t) \\
\epsilon_k (t) && g X_b(t)
\end{pmatrix},
\\
\hat \Sigma(t,t') &=
\begin{pmatrix}
\Sigma_a (t,t') && 0 \\
0 && \Sigma_b(t,t')
\end{pmatrix},
\end{align}
with the sublattice-dependent $\Sigma_{a,b}$ and $\langle X\rangle_{a,b}$.

\section{Derivation the reflectivity fluctuations}

In this section we present explicit steps of the fluctuation expansion leading from the general expression for the moments $I_n$ of the photon-count [main text, Eq.~(2)]  to the variance of the intensity [main text, Eq.~(1)]. We start from the expression for $I_n$,which was been derived by Fleischhauer \cite{Fleischhauer1998,footnote},
\begin{align}
I_1
=&
\int d1 d1' d\bar 1 d\bar 1' \,\,\mathcal{D}_{11'}
g(1,\bar 1) g(1',\bar 1')\langle j(\bar 1) j(\bar 1')\rangle,
\label{I1app}
\\
I_2
=&
\int d1 d1' d2 d2' d\bar 1 d\bar 1' d\bar 2 d\bar 2'\,\,\mathcal{D}_{11'}\mathcal{D}_{22'}
g(1,\bar 1) g(1',\bar 1')
\nonumber\\&
\times\,\,\,\,g(2,\bar 2) g(2',\bar 2')
\langle
T_{\bar \tau}[j(\bar 1)j(\bar 2)]
T_{\tau}[j(\bar 1')j(\bar 2')]
\rangle.
\label{I2app}
\end{align}
Here 
\begin{align}
\label{Ddef}
\mathcal{D}_{1,1'}=\epsilon \delta(\bm r_1-\bm R)\delta(\bm r_1'-\bm R) \int_{0}^\infty \! d\omega \,e^{-i\omega (t_1-t_1')}
\end{align}
is the detector response function, and $g$ is the linear kernel which relates the induced field $E_{ind}$ and the current $j$ by a solution of Maxwell equations, 
\begin{align}
\label{maxwelll}
E_{ind}(1)=\int d\bar 1 g(1,\bar 1) j(\bar 1).
\end{align}

In Eq.~\eqref{I1app} and \eqref{I2app}, we insert the expansion $j(1)=\langle j(1)\rangle+ \delta j(1)$. Terms which are first order in $\delta j$ vanish by construction, because $\langle\delta j\rangle=0$. Zeroth order terms are identical in $I_2$ and $I_1^2$, and thus vanish in the variance $I_2-I_1^2$. Third and fourth order terms, such as Eq.~\eqref{I2app} where the correlation function in the integrand is replaced by $\langle T_{\bar \tau}[\delta j(\bar 1) \delta j(\bar 2)] \delta j(\bar 1')\rangle\langle j(\bar 2') \rangle$, are not considered here as explained in the main text, because they scale differently with the sample volume. To second order in $\delta j$, Eq.~\eqref{I2app} has six terms, where the  current correlation function in the integral takes one of the following combinations
\begin{align}
&\langle  T_{\bar \tau}[\delta j(\bar 1)\delta j(\bar 2)] \rangle\langle j(\bar 1')\rangle \langle j(\bar 2') \rangle,
\,\,%\nonumber\\&+
 \langle  \delta j(\bar 1)\delta j(\bar 2')\rangle\langle j(\bar 1')\rangle \langle j(\bar 2) \rangle,
\nonumber\\&%+
\langle  \delta j(\bar 2) \delta j(\bar 1') \rangle\langle j(\bar 1)\rangle \langle j(\bar 2') \rangle,
\,\,%\nonumber\\&%+
\langle T_{\tau}[ \delta j(\bar 1')\delta j(\bar 2') ]\rangle\langle j(\bar 1)\rangle \langle j(\bar 2) \rangle,
\nonumber\\&%+
\langle  \delta j(\bar 1)\delta j(\bar 1') \rangle\langle j(\bar 2)\rangle \langle j(\bar 2') \rangle,
\,\,%\nonumber\\&%+
\langle  \delta j(\bar 2)\delta j(\bar 2') \rangle\langle j(\bar 1)\rangle \langle j(\bar 1') \rangle.
\label{sixterms}
\end{align}
Here the time-ordering operator can be dropped whenever it acts on c-numbers $\langle j \rangle$, such as for the second term,
$\langle T_{\bar \tau} [\delta j(\bar 1)\langle j(\bar 2) \rangle]T_{\tau}[\delta j(\bar 2')\langle j(\bar 1')\rangle] \rangle=\langle  \delta j(\bar 1')\delta j(\bar 2) \rangle\langle j(\bar 1)\rangle \langle j(\bar 2') \rangle$. Of the six terms in Eq.~\eqref{sixterms}, the last two are cancelled by corresponding terms in the expansion of $I_1^2$. For the remaining four, one can evaluate integrals in \eqref{I2app} which correspond to a contraction of the current expectation values $\langle j \rangle$ with $\mathcal{D}$,
\begin{align}
\nonumber
&\int d\bar 1 d1\,\,
 \mathcal{D}_{1,1'} g(1,\bar 1) \langle j(\bar 1)\rangle
 \\
 &=
\epsilon \delta(\bm r'_1-\bm R) \int d1
\,E_{ind}(1) \,\delta(\bm r_1-\bm R) \int_0^\infty\!\!\!\!\!d\omega \,e^{-i\omega(t_1-t_1')}
 \nonumber\\&=
\delta(\bm r_1'-\bm R) \int_0^\infty\!\!\!\!\!d\omega \,e^{-i\omega(t_1-t_1')} E_{ind}(\bm R,t_1)
\nonumber
\\
&
\equiv
\epsilon \delta(\bm r_1'-\bm R) E_-(t_1'),
\label{eminus}
\end{align}
using Eqs.~\eqref{Ddef} and \eqref{maxwelll} in the first step. Similarly,
\begin{align}
\nonumber
&\int d\bar 1' d1'\,\,
 \mathcal{D}_{1,1'} g(1',\bar 1') \langle j(\bar 1')\rangle
 \\
 &=
\epsilon \delta(\bm r_1-\bm R) \int d1
\,E_{ind}(1') \,\delta(\bm r_1'-\bm R) \int_0^\infty\!\!\!\!\!d\omega \,e^{-i\omega(t_1-t_1')}
 \nonumber
 \\
 &
 =
\delta(\bm r_1-\bm R) \int_0^\infty\!\!\!\!\!d\omega \,e^{-i\omega(t_1-t_1')} E_{ind}(\bm R,t_1')
\nonumber
\\
&
\equiv
\epsilon \delta(\bm r_1-\bm R) E_+(t),
\label{eplus}
\end{align}
Inserting Eqs.~\eqref{sixterms}, \eqref{eminus}, and \eqref{eplus} into Eqs.~\eqref{I2app} and \eqref{I1app} we get
\begin{align}
&I_2-I_1^2=
%\int d1 d1' d2 d2' d\bar 1 d\bar 1' d\bar 2 d\bar 2'\,\,\mathcal{D}_{11'}\mathcal{D}_{22'}
%g(1,\bar 1) g(1',\bar 1')
\nonumber\\
=&\epsilon^2
\int d1 d2 d\bar 1 d\bar 2 \,g(1,\bar 1)g(2,\bar 2)\,\,\,\times
\nonumber\\&\times
\langle  T_{\bar \tau}[\delta j(\bar 1)\delta j(\bar 2)] \rangle  \delta(\bm r_1-\bm R) E_+(t_1) \delta(\bm r_2-\bm R) E_+(t_2)
\nonumber\\&+
\epsilon^2
\int d1 d2' d\bar 1 d\bar 2' \,g(1,\bar 1)g(2',\bar 2')\,\,\,\times
\nonumber\\&\times
\langle  \delta j(\bar 1)\delta j(\bar 2') \rangle
\delta(\bm r_1-\bm R) E_+(t_1) \delta(\bm r_2'-\bm R) E_-(t_2')
\nonumber\\&+
\epsilon^2
\int d1' d2 d\bar 1' d\bar 2 \,g(1',\bar 1')g(2,\bar 2)\,\,\,\times
\nonumber\\&\times
\langle  \delta j(\bar 2)\delta j(\bar 1') \rangle
\delta(\bm r_1'-\bm R) E_-(t_1') \delta(\bm r_2-\bm R) E_+(t_2)
\nonumber\\&+
\epsilon^2
\int d1' d2' d\bar 1' d\bar 2'\, g(1',\bar 1')g(2',\bar 2')\,\,\,\times
\nonumber\\&\times
\langle T_{\tau}[ \delta j(\bar 1')\delta j(\bar 2') ]\rangle
\delta(\bm r_1'-\bm R) E_-(t_1') \delta(\bm r_2'-\bm R) E_-(t_2').
\nonumber
\end{align}
%%%%%%%%%%%%%%%%%%%%%%%%%%%
\begin{align}
=&\epsilon^2
\int dt_1 dt_2 
\langle  T_{\bar \tau}[\delta j( \bm R,t_1)\delta j( \bm R,t_2)] \rangle_{ret}  
 E_+(t_1)  E_+(t_2)
\nonumber\\&+
\epsilon^2
\int dt_1 dt_2' 
\langle  \delta j( \bm R,t_1)\delta j(\bm R,t_2') \rangle_{ret}
 E_+(t_1)  E_-(t_2')
\nonumber\\&+
\epsilon^2
\int dt_1' dt_2 
\langle  \delta j( \bm R,t_2)\delta j(\bm R,t_1') \rangle_{ret}
 E_-(t_1') E_+(t_2)
\nonumber\\&+
\epsilon^2
\int dt_1' dt_2' 
\langle T_{\tau}[ \delta j(\bm R,t_1')\delta j(\bm R, t_2') ]\rangle_{ret}
E_-(t_1') E_-(t_2').
\nonumber
%%%%%%%%%%%%%%%%%%%%%%%%%%%%%%%%%%%%%%%%%%%
\\
=&2\epsilon^2 \text{Re}
\int dt dt' 
\langle  T_{\bar \tau}[\delta j( \bm R,t)\delta j( \bm R,t')] \rangle_{ret}  
 E_+(t)  E_+(t')
\nonumber\\&+
2\epsilon^2
\int dt dt' 
\langle  \delta j( \bm R,t)\delta j(\bm R,t') \rangle_{ret}
 E_+(t)  E_-(t').
\end{align}
In the main text the first term is not discussed because it  would vanish by averaging over a carrier envelope phase $\varphi$ ($E_{\pm}\sim e^{\mp i\varphi}$). The remaining term is Eq.~(1) of the main text.

\end{appendix}


\begin{thebibliography}{100}
\newcommand{\bibtitle}[1]{``#1''}

\bibitem{Giannetti2016}
C.~Giannetti, M.~Capone, D.~Fausti, M.~Fabrizio, F.~Parmigiani, and D.~Mihailovic,
\bibtitle{Ultrafast optical spectroscopy of strongly correlated materials and high-temperature superconductors: a non-equilibrium approach}, 
%Adv. Phys. {\bf 65} (2), 58-238 (2016).
Adv. Phys. {\bf 65}, 58 (2016).

\bibitem{Esposito2015}
M.~Esposito, K.~Titimbo, K.~Zimmermann, F.~Giusti, F.~Randi, D.~Boschetto, F.~Parmigiani, R.~Floreanini, F.~Benatti, and D.~Fausti,
\bibtitle{Photon number statistics uncover the fluctuations in non-equilibrium lattice dynamics},
Nat.~Comm. {\bf 6}, 10249 (2015).

\bibitem{Tesi}
F.~Randi,
\bibtitle{Pulsed homodyne detection for quantum state reconstruction applied to ultrafast non-equilibrium spectroscopy},
Master thesis, Universit\`{a} degli Studi di Trieste (2011).

\bibitem{Misochko2001}
O.V.~Misochko,
\bibtitle{Coherent Phonons and Their Properties},
Journal of Experimental and Theoretical Physics, {\bf 92}, 246 (2001).
%Journal of Experimental and Theoretical Physics, {\bf 92}(2):246--259 (2001).

\bibitem{Riek2017}
C.~Riek, P.~Sulzer, M.~Seeger, A. S.~Moskalenko, G.~Burkard, D. V.~Seletskiy, and A.~Leitenstorfer,
\bibtitle{Subcycle quantum electrodynamics},
Nature {\bf 541}, 376, (2017).

\bibitem{Beenakker2001}
C. W. J. Beenakker and H. Schomerus
\bibtitle{Counting Statistics of Photons Produced by Electronic Shot Noise},
Phys. Rev. Lett. {\bf 86}, 700 (2001).
%DOI: https://doi.org/10.1103/PhysRevLett.86.700

\bibitem{Lebedev2010}
A. V. Lebedev, G. B. Lesovik, and G. Blatter
\bibtitle{Statistics of radiation emitted from a quantum point contact},
Phys. Rev. B {\bf 81}, 155421 (2010).
%https://doi.org/10.1103/PhysRevB.81.155421

\bibitem{Benatti2017}
F.~Benatti, M.~Esposito, D.~Fausti, R.~Floreanini, K.~Titimbo, and K.~Zimmermann,
\bibtitle{Generation and detection of squeezed phonons in lattice dynamics by ultrafast optical excitations},
N. J. Phys. {\bf 19}, 023032 (2017).

\bibitem{KelvinThesis}
K. Titimbo,
\bibtitle{Creation and detection of squeezed phonons in pump and probe experiments: a fully quantum treatment},
PhD Thesis, Universit\`{a} degli Studi di Trieste (2011).


\bibitem{Fleischhauer1998}
M.~Fleischhauer,
\bibtitle{Quantum-theory of photodetection without the rotating wave approximation},
J. Phys. A: Math. Gen. {\bf 31},  453 (1998).

\bibitem{Kelley1964}
P.L. Kelley and W.H. Kleiner,
\bibtitle{Theory of Electromagnetic Field Measurement and Photoelectron Counting},
Phys. Rev. {\bf 136}, A316 (1964).

\bibitem{Mandel1958}
L. Mandel, 
\bibtitle{Fluctuations of Photon Beams and their Correlations},
Proc. Phys. Soc. {\bf 72}, 1037 (1958).



\bibitem{footnote1}
Counter-rotating terms $\sim E_+ E_+$ have been omitted, as they vanish in a typical pump-probe experiment which averages over the carrier envelope phase of the probe $E_+\propto e^{i\phi}$.

%\bibitem{Supplemental}
%ME
%See supplemental material for details of the numerical implementation of the non-equilibrium DMFT formalism, and details of the derivation of Eq.~\eqref{eqI33}.

%%%%%%%%%%

\bibitem{Eckstein2008}
M.~Eckstein, and M.~Kollar
\bibtitle{Theory of time-resolved optical spectroscopy on correlated electron systems},
Phys. Rev. B {\bf 78}, 205119 (2008).

\bibitem{Kubo1957}
Kubo, R., 
\bibtitle{Statistical-Mechanical Theory of Irreversible Processes. I. General Theory and Simple Applications to Magnetic and Conduction Problems},
J. Phys. Soc. Jpn. {\bf 12}, 570 (1957).

\bibitem{Peierls1991}
R.~E.~Peierls,
\bibtitle{More surprises in theoretical physics}
Princeton University Press (1991).

\bibitem{NonEqDMFT}
H.~Aoki, N.~Tsuji, M.~Eckstein, M.~Kollar, T.~Oka, and P.~Werner,
\bibtitle{Nonequilibrium dynamical mean-field theory and its applications},
Rev. Mod. Phys. {\bf 86}, 779 (2014).

\bibitem{Murakami2015}
Y.~Murakami, P.~Werner, N.~Tsuji, and H.~Aoki,
\bibtitle{Interaction quench in the Holstein model: Thermalization crossover from electron- to phonon-dominated relaxation},
Phys. Rev. B {\bf 91}, 045128 (2015).

\bibitem{Teitelbaum2016}
S.~W.~Teitelbaum, T.~Shin, J.~Wolfson, Y.~H.~Cheng, I.~J.~Porter, M.~Kandyla, and K.~A.~Nelson
\bibtitle{A Photoinduced Symmetric Crystalline Phase in Bismuth},
arXiv:1609.04048.

\bibitem{Kogoj2016}
J.~Kogoj, L.~Vidmar, M.~Mierzejewski, S.A.~Trugman, and J.~Bon\v{c}a,
\bibtitle{Thermalization after photoexcitation from the perspective of optical spectroscopy},
Phys. Rev. B {\bf 94}, 014304 (2016). 

\bibitem{footnote02}
Note that because $\delta t$ is considerably smaller than one oscillation period, and also the corresponding frequency uncertainty is small compared to $\omega_p$, the result approximately corresponds to a measurement of $C(t_p,\omega_p)$ itself,
%ME01
independent of the precise pulse shape.

\bibitem{footnote}
Note that we use a different notation with respect to Fleischhauer~\cite{Fleischhauer1998}. In particular, our kernel $g(1,2)$ corresponds to Fleischhauer's $D^\mathrm{ret}(1,2)$, our detector response $\mathcal{D}_{1,2}$ corresponds to Fleischhauer's $f(t_1,t_2)$, and our currents $j$ are Fleischhauer's sources $s$.


\end{thebibliography}
\end{document}